\documentclass[english,aps,preprint]{revtex4}
\usepackage[]{fontenc}
\usepackage[latin9]{inputenc}
\usepackage{graphicx}
\usepackage{amssymb}
\usepackage{babel}

\begin{document}

\preprint{Submitted to Phys. Rev. E}

\title{Dimensional phase transitions in small Yukawa clusters}

\author{T. E. Sheridan}

\email{t-sheridan@onu.edu}

\author{K. D. Wells}

\affiliation{Department of Physics \& Astronomy, Ohio Northern University, Ada,
OH 45810}

\begin{abstract}
We investigate the one- to two-dimensional zigzag transition in clusters
consisting of a small number of particles interacting through a Yukawa
(Debye) potential and confined in a two-dimensional biharmonic potential
well. Dusty (complex) plasma clusters with $n\le19$ monodisperse
particles are characterized experimentally for two different confining
wells. The well anisotropy is accurately measured, and the Debye shielding
parameter is determined from the longitudinal breathing frequency.
Debye shielding is shown to be important. A model for this system
is used to predict equilibrium particle configurations. The experiment
and model exhibit excellent agreement. The critical value of $n$
for the zigzag transition is found to be less than that predicted
for an unshielded Coulomb interaction. The zigzag transition is shown
to behave as a continuous phase transition from a one-dimensional
to a two-dimensional state, where the state variables are the number
of particles, the well anisotropy and the Debye shielding parameter.
A universal critical exponent for the zigzag transition is identified
for transitions caused by varying the Debye shielding parameter.
\end{abstract}


\maketitle

\section{Introduction}

Consider a strongly-coupled, two-dimensional (2D) system of $n$ particles
with identical mass $m$ and charge $q$ . A confining potential well
is required to balance the repulsive interparticle force and create
a stable configuration. For almost any 2D potential well expanded
around its minimum, the lowest order terms in a particle's potential
energy are

\begin{equation}
U(x,y)\approx U_{0}+\left.\frac{1}{2}\frac{\partial^{2}U}{\partial x^{2}}\right|_{0}x^{2}+\left.\frac{1}{2}\frac{\partial^{2}U}{\partial y^{2}}\right|_{0}y^{2},\label{eq:Vxy}\end{equation}
where $U_{0}$ is a constant. Consequently, we can approximate the
confining potential energy as\begin{equation}
U\left(x,y\right)=\frac{1}{2}k_{x}x^{2}+\frac{1}{2}k_{y}y^{2}=\frac{1}{2}m\omega_{0x}^{2}x^{2}+\frac{1}{2}m\omega_{0y}^{2}y^{2},\label{eq:biharm}\end{equation}
where $k_{x}$ and $k_{y}$ are force constants, and $\omega_{0x}$
and $\omega_{0y}$ are single-particle (center-of-mass) oscillation
frequencies in the $x$ and $y$ directions, respectively. In 2D,
the biharmonic well {[}Eq. (\ref{eq:biharm})] gives the general confining
potential energy when higher order terms are negligible. 

When charged particles are in free space, they interact through an
unshielded Coulomb potential \cite{sch,dev,bed}. However, if the
particles are in a dielectric, then the Coulomb interaction is shielded
by the medium's dielectric response, and particles interact through
a Yukawa potential (i.e., a shielded Coulomb or Debye potential),

\begin{equation}
V(r)=\frac{1}{4\pi\epsilon_{0}}\frac{q}{r}e^{-r/\lambda},\label{eq:debye}\end{equation}
where $r$ is the separation distance, and $\lambda$ is the Debye
length. We call a system of particles confined to two dimensions and
interacting through a Yukawa potential a {}``2D Yukawa system''.
If $n$ is small, then the system is a {}``Yukawa cluster''. In
2D Yukawa systems the finite Debye length allows the particle-particle
interaction length to be varied from long range to short range \cite{tes7},
affecting both the system's static and dynamic properties.

The case of 2D Yukawa systems in isotropic ($\omega_{0x}^{2}=\omega_{0y}^{2}$)
potential wells has been explored extensively, both theoretically
\cite{kon,lai,tes5} and experimentally \cite{juan,tes2,tes3,ivan,mel2}.
For isotropic wells, large-$n$ systems form a circular disk where
the interior of the disk has a triangular lattice \cite{bed}. For
clusters (small-$n$ systems), different shell configurations become
stable as the interaction is tuned from long range to short range.
When the potential well is anisotropic, qualitatively new types of
configurations can occur \cite{sch,can,apo}. If the well is weakly
anisotropic then clusters are elliptical and have well-defined shell
structures \cite{can,tes4}. On the other hand, when the well is highly
anisotropic, the particle configuration is a one-dimensional (1D)
straight line \cite{hom,mis,liu,pia}. A 1D cluster becomes a 2D cluster
through a zigzag transition. Zigzag configurations may become elliptical
and then circular through further structural transitions \cite{can}.

Dusty (complex) plasma should be an ideal experimental system for
studying the zigzag transition in 2D Yukawa clusters. In laboratory
dusty plasmas, monodisperse dust particles interacting through a Yukawa
potential \cite{ukon} are confined near the sheath edge above a horizontal
electrode to form a 2D system. A rectangular confining structure placed
on top of the electrode can produce a biharmonic potential well \cite{hom,mel1,tes4}. 

Melzer \cite{mel1} experimentally observed zigzag transitions in
dusty plasmas confined in a radio frequency (rf) discharge as a function
of particle number $n$ and neutral gas pressure, and attempted to
infer the well anisotropy and Debye shielding parameter using a static
analysis of the cluster configuration together with a comparison to
unshielded Coulomb theory. He concluded that the measured cluster
properties, including the critical value of $n$ for the zigzag transition,
were not inconsistent with the physics of an unshielded Coulomb interaction
(i.e., $\lambda\rightarrow\infty$).

In this paper, we study Yukawa clusters in one- and two-dimensional
configurations and the transition between these configurations. Dusty
plasma experiments are performed as a function of particle number
for two rectangular confining wells, giving two values of the well
anisotropy. This work extends previous experiments \cite{tes4} on
2D Yukawa clusters in weakly anisotropic wells. We directly measure
the well anisotropy and Debye shielding parameter \cite{tes4}. We
find that Debye shielding is important, i.e., our results are \emph{not}
consistent with physics in the unshielded Coulomb regime. Using the
measured cluster parameters, we compute predicted equilibrium configurations
from the model of Sec. \ref{sec:Model}. The predicted and measured
configurations exhibit excellent agreement. The critical value of
$n$ for the zigzag transition is found to be the same in both experiment
and theory, and to be less than that predicted for an unshielded Coulomb
interaction. Even though $n$ is small, the zigzag transition is shown
to behave as a 1D-2D continuous phase transition and a universal critical
exponent is identified.

\section{\label{sec:Model}Model}

Two-dimensional Yukawa clusters can be modeled as a strongly-coupled
system of $n$ identical particles with charge $q$ and mass $m$
at positions $\left\{ x_{i},y_{i}\right\} $ interacting through a
Yukawa potential {[}Eq. (\ref{eq:debye})] with Debye length $\lambda$.
The particles are confined in a 2D biharmonic well {[}Eq. (\ref{eq:biharm})]
where $\omega_{0x}$ and $\omega_{0y}$ are oscillation frequencies
for the $x$ (longitudinal) and $y$ (transverse) directions, respectively.
The separation distance between particles $i$ and $j$ is $r_{ij}=\sqrt{\left(x_{i}-x_{j}\right)^{2}+\left(y_{i}-y_{j}\right)^{2}}$.
The total potential energy of the system is \cite{can,tes4} \begin{equation}
U=\sum_{i=1}^{n}\left(\frac{1}{2}m\omega_{0x}^{2}x_{i}^{2}+\frac{1}{2}m\omega_{0y}^{2}y_{i}^{2}\right)+\sum_{j>i=1}^{n}\left(\frac{q^{2}}{4\pi\epsilon_{0}}\frac{e^{-r_{ij}/\lambda}}{r_{ij}}\right),\label{eq:potential}\end{equation}
where the first sum in Eq. (\ref{eq:potential}) is the potential
energy of confinement, and the second sum is the potential energy
due to particle-particle interactions. Equation (\ref{eq:potential})
can be nondimensionalized to give \begin{equation}
\frac{U}{U_{0}}=\sum_{i=1}^{n}\left(\xi_{i}^{2}+\alpha^{2}\eta_{i}^{2}\right)+\sum_{j>i=1}^{n}\left(\frac{e^{-\kappa\rho_{ij}}}{\rho_{ij}}\right),\label{eq:nondim_pot}\end{equation}
where $U_{0}$ is the characteristic potential energy, and $\xi_{i}=x_{i}/r_{0}$,
$\eta_{i}=y_{i}/r_{0}$, and $\rho_{ij}=r_{ij}/r_{0}$ are normalized
distances. We define the characteristic length scale \begin{equation}
r_{0}^{3}=\frac{2}{m\omega_{0x}^{2}}\frac{q^{2}}{4\pi\epsilon_{0}}\label{eq:nondim_para}\end{equation}
using the longitudinal oscillation frequency $\omega_{0x}$. The dimensionless
parameters in Eq. (\ref{eq:nondim_pot}) are the particle number $n$,
the well anisotropy $\alpha^{2}$ and the Debye shielding parameters
$\kappa$, where \begin{equation}
\alpha^{2}=\frac{k_{y}}{k_{x}}=\frac{\omega_{0y}^{2}}{\omega_{0x}^{2}},\;\kappa=\frac{r_{0}}{\lambda},\label{eq:ka2}\end{equation}
respectively. To compare this model to experiment, $\alpha^{2}$ and
$\kappa$ must be measured in the experiment.

This model {[}Eq. (\ref{eq:nondim_pot})] has three parameters: $n$,
$\alpha^{2}$ and $\kappa$. The isotropic well is given by $\alpha^{2}=1$.
Without loss of generality, we assume that the anisotropic well has
$\alpha^{2}>1$ (i.e., $\omega_{0y}^{2}>\omega_{0x}^{2}$) so that
the major axis of the potential well lies in the $x$ (longitudinal)
direction. An unshielded Coulomb interaction corresponds to $\kappa=0$.
As $\kappa$ increases the interparticle force becomes more localized.
Given $n$, $\alpha^{2}$ and $\kappa$, a solution of the model {[}Eq.
(\ref{eq:nondim_pot})] is a set of particle positions $\left\{ \xi_{i},\eta_{i}\right\} $
that minimizes $U$. Properties of such solutions have previously
been investigated by C\^andido, et al. \cite{can}. To minimize $U/U_{0}$
we use simulated annealing together with a final step of conjugate
gradient minimization \cite{tes5}. For a given configuration, normal
modes and their associated frequencies can be computed from the dynamical
matrix.

A zigzag transition \cite{sch} is a transition from a 1D straight
line configuration to a 2D configuration. For particle coordinates
$\left\{ x_{i},y_{i}\right\} $ measured with respect to the cluster's
center of mass, the cluster's length and width can be characterized
by the rms values\begin{equation}
x_{rms}^{}=\sqrt{\frac{1}{n}\sum x_{i}^{2}},\; y_{rms}^{}=\sqrt{\frac{1}{n}\sum y_{i}^{2}}.\label{eq:yrms}\end{equation}
Consequently, a zigzag transition is a transition from $y_{rms}=0$
to $y_{rms}>0$ caused by a change in one of the model parameters.
If a cluster is initially in a straight line configuration, then for
constant $\alpha^{2}$ and $\kappa$ a zigzag transition will occur
as $n$ is increased. We denote the critical value of $n$, which
is the smallest value of $n$ in the zigzag configuration, by $n_{c}$.
A 1D-2D transition also occurs when $\kappa$ is increased above a
critical value $\kappa_{c}$. If a cluster is initially in a 2D configuration,
then increasing $\alpha^{2}$ causes a transition to a 1D cluster
\cite{can} for which $y_{i}=\eta_{i}=0$ above the critical value
$\alpha_{c}^{2}$. As a consequence, 1D configurations are independent
of $\alpha^{2}$ when $\alpha^{2}>\alpha_{c}^{2}$.

An \emph{unbounded} 1D chain can be modeled by letting $\omega_{0x}\rightarrow0$
while $\omega_{0y}$ remains finite. Longitudinal confinement can
be achieved either by using periodic boundary conditions \cite{pia}
or a ring topology \cite{tesp}. It is then convenient to define the
characteristic length scale using the transverse frequency $\omega_{0y}$
\cite{pia,mel1}, \begin{equation}
r_{0T}^{3}=\frac{2}{m\omega_{0y}^{2}}\,\frac{q^{2}}{4\pi\epsilon_{0}}=\frac{r_{0}^{3}}{\alpha^{2}}.\label{eq:rot}\end{equation}
This gives a transverse Debye shielding parameter $\kappa_{T}$ which
is related to $\kappa$ {[}Eq. (\ref{eq:ka2})] by \begin{equation}
\kappa_{T}=\frac{r_{0T}}{\lambda}=\frac{\kappa}{\left(\alpha^{2}\right)^{1/3}}\le\kappa.\label{eq:kt}\end{equation}
A zigzag transition occurs when the 1D lattice constant $a<a_{c}$
where $a_{c}$ is a critical value \cite{pia,tesp}. For an unbounded
Yukawa chain, the dimensionless critical lattice constant $a_{c}/r_{0T}$
is a solution of \cite{tesp}\begin{equation}
\left(\frac{a_{c}}{r_{0T}}\right)^{3}=2\sum_{j=1,3,...}\frac{e^{-j\kappa_{T}\left(a_{c}/r_{0T}\right)}}{j^{3}}\left(1+j\kappa_{T}\frac{a_{c}}{r_{0T}}\right),\label{eq:ac}\end{equation}
which depends only on the transverse shielding parameter $\kappa_{T}$.
For a pure Coulomb interaction, $\kappa_{T}=0$, $\left(a_{c}/r_{0T}\right)^{3}=\left(7/4\right)\zeta\left(3\right)$
so that $a_{c}/r_{0T}\approx1.28$. The critical lattice spacing decreases
as the Debye shielding parameter increases (see Fig. \ref{fig:acrit}
below).

\section{Experiment}

\begin{figure}
\includegraphics[width=3.25in]{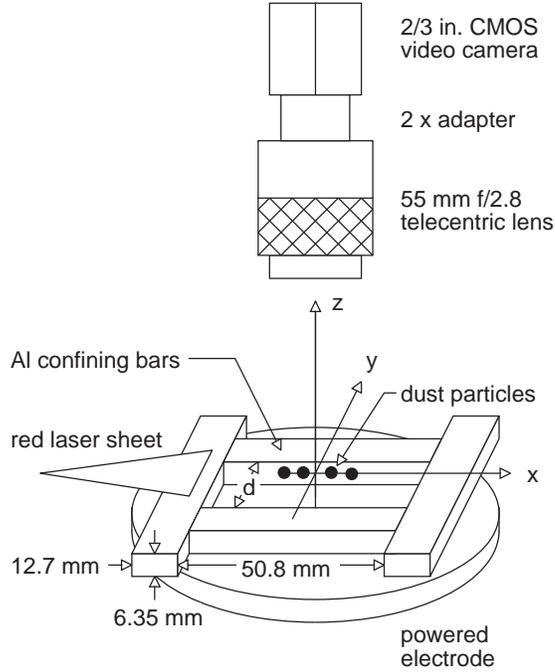}

\caption{\label{fig:scheme}Schematic of the experimental setup. Nearly identical
spherical dust particles are confined in a biharmonic potential well
created in the rectangular depression between four conducting bars
placed on the rf powered electrode. Experiments were performed for
confinement geometries with bar separations $d=25.4\;{\rm mm}$ and
$d=14.0\;{\rm mm}$. }

\end{figure}

Dusty plasma experiments were performed in the Dusty Ohio Northern
University experimenT (DONUT) \cite{tes1,tes2,tes3,tes4,tes6}. An
argon plasma was created around an 89-mm diameter powered electrode
in a radio frequency discharge at 13.56 MHz. A blocking capacitor
allows the electrode to develop a negative dc self-bias that levitates
the negatively-charged dust particles. As shown in Fig. \ref{fig:scheme},
the biharmonic well is formed at the minimum of a confining geometry
consisting of four rectangular aluminum bars placed on the powered
electrode. The end bars measure 6.35 mm $\times$ 12.7 mm $\times$
76.2 mm, while the inner bars measure 6.35 mm $\times$ 12.7 mm $\times$
50.8 mm. The distance $d$ between the two inner bars can be changed
to vary the dimensions of the confining rectangular depression, and
thereby change the anisotropy parameter $\alpha^{2}$. Clusters were
made using monodisperse melamine formaldehyde spheres with a nominal
diameter of $9.62\pm0.09\,{\rm \mu m}$. As explained previously \cite{tes2},
we believe that the dust particle diameter is closer to $8.94\pm0.18\;{\rm \mu m}$.

To determine dust particle positions, the particles are illuminated
by a red diode laser and viewed using a $2/3$ inch CMOS camera with
a telecentric lens mounted above the top face of the electrode. For
these experiments, we recorded 4097 frames of video at $\approx30$
frames/s for each particle configuration to determine center-of-mass
(c.m.) and breathing frequencies. A side-view camera was used to verify
that out-of-plane motion was minimal.

Two different confinement geometries were studied for similar plasma
conditions. In the first, the inner bars were separated by $d=25.4\;{\rm mm}$,
while in the second they were separated by $d=14.0\;{\rm mm}$. For
the 25.4 mm $\times$ 50.8 mm well, the neutral Ar pressure was 12.4
mtorr (1.65 Pa), the rf power was $\approx10\;{\rm W}$ forward, the
dc self bias on the electrode was $-$89.0 V, and particle positions
were recorded with a resolution of 16.51 $\mu{\rm m/pixel}$. For
the 14.0 mm $\times$ 50.8 mm well, the neutral Ar pressure was 12.1
mtorr (1.61 Pa), the rf power was $\approx9\;{\rm W}$ forward, the
dc self bias was $-83.0\;{\rm V}$, and positions were recorded with
a resolution of 16.77 $\mu{\rm m/pixel}$.

Normal mode frequencies were determined by projecting the particle's
thermal motion onto the center-of-mass and longitudinal breathing
modes \cite{tes1,tes3}. A Fourier transform of the time history of
the mode amplitude gives the power spectral density for that mode,
which is that of a driven damped harmonic oscillator. For the neutral
pressures used, the oscillations are underdamped and the power spectra
display a clear resonance peak. Measuring the center-of-mass frequencies
$\omega_{0x}$ and $\omega_{0y}$ directly determines the anisotropy
parameter $\alpha^{2}$, while comparing the longitudinal breathing
frequency for 1D configurations to model solutions determines the
Debye shielding parameter $\kappa$.

\section{\label{sec:Results}Experimental Results}

\begin{figure}
\includegraphics[width=3.25in]{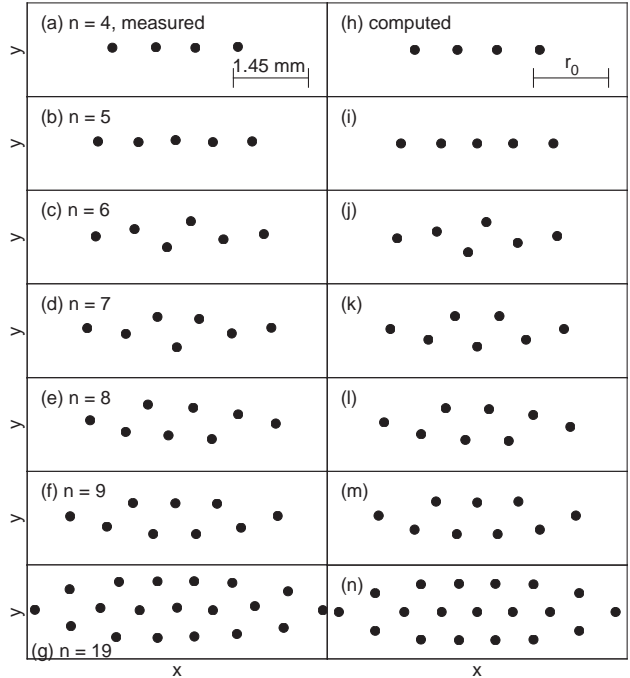}

\caption{\label{fig:d25pos}(a)-(g) Measured particle positions for confining
bar separation $d=25.4$ mm. (h)-(n) Computed positions for $\alpha^{2}=9.24$
and $\kappa=3$. By matching $y_{rms}$ between the experiment and
model, we find $r_{0}\approx1.40\;{\rm mm}$. Agreement between measured
and computed configurations is excellent. For $n\le5$ the configurations
are linear, at $n=6$ a zigzag develops and for $n=19$ a fully elliptical
cluster with a well-defined shell structure is seen. Both measured
and computed figures have a 1:1 aspect ratio.}

\end{figure}

For the 25.4 mm $\times$ 50.8 mm confining well, nine sets of particles
were analyzed with $n=2$ to 19. Representative configurations are
shown in Figs. \ref{fig:d25pos}(a)-(g). For $n\le5$ the particles
are in a 1D linear configuration. When one more particle is added
($n=6$) the cluster changes to a 2D zigzag configuration, so $n_{c}=6$.
As $n$ increases the number of zigzags also increases until zigzags
stretch from one end of the cluster to the other ($n=8$, 9). For
$n=19$ the system displays a full (5,14) elliptical shell structure
\cite{can,tes4}. 

\begin{figure}
\includegraphics[width=3.25in]{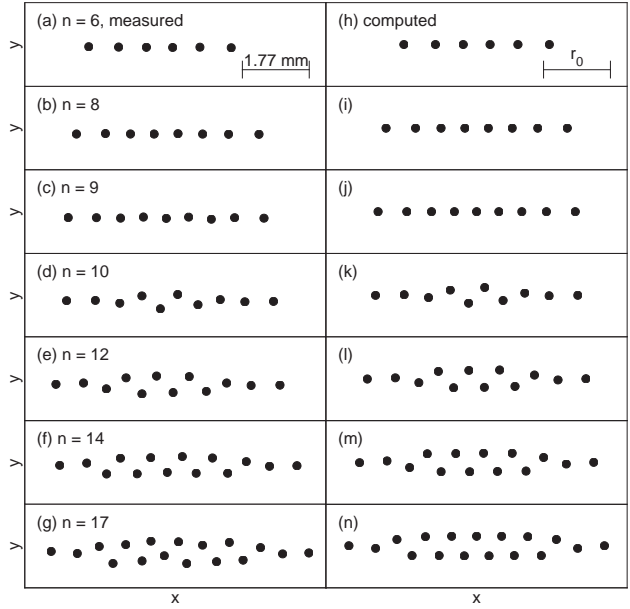}

\caption{\label{fig:d14pos}(a)-(g) Measured particle positions for a confining
well with $d=14.0\;{\rm mm}$. (h)-(n) Computed positions for $\alpha^{2}=30.7$
and $\kappa=4$. We estimate $r_{0}\approx1.65\;{\rm mm}$ by comparing
$y_{rms}$ for the model and experiment. For $n\le9$ the configurations
are linear, at $n=10$ a zigzag develops. For $n=17$ the cluster
remains in a zigzag configuration. Both measured and computed figures
have a 1:1 aspect ratio.}

\end{figure}

For the narrower confining well, $d=14.0\;{\rm mm}$, we analyzed
twelve sets of particles for nine different values of $n$ ($2\le n\le17$).
Measured configurations are shown in Figs. \ref{fig:d14pos}(a)-(g).
In comparison to $d=25.4\;{\rm mm}$, we expect the anisotropy parameter
to be larger so that the critical value of $n$ is increased. For
these conditions, clusters with $n\le9$ are in a 1D configuration.
A zigzag configuration is seen for $n=10$, so $n_{c}=10$. As $n$
further increases the zigzag region expands away from the center of
the cluster. However, even for $n=17$ the cluster still has short
linear tails at each end \cite{can} and is not an elliptical configuration.

\begin{figure}
\includegraphics[width=3.25in]{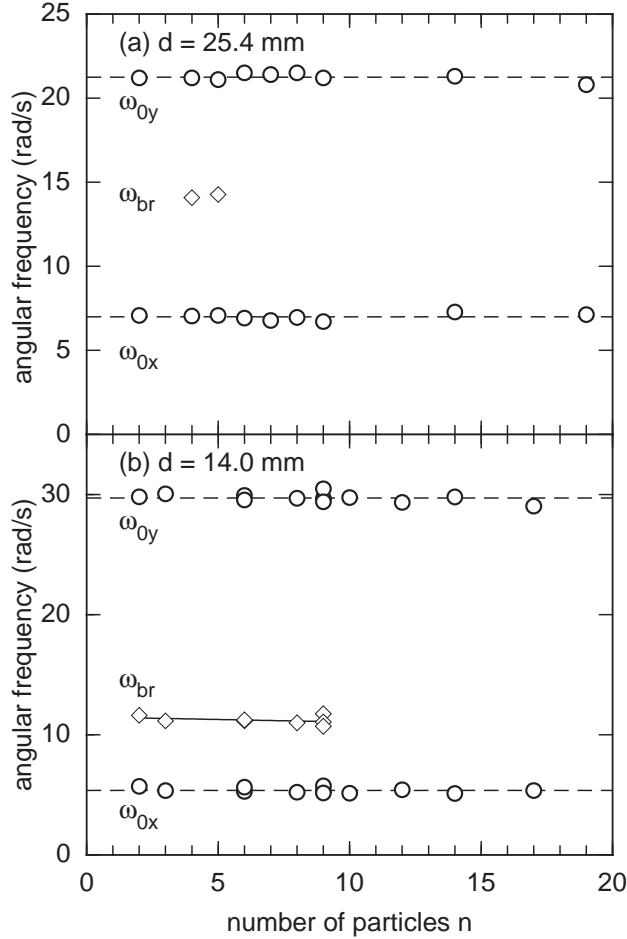}

\caption{\label{fig:wxwy}Measured center-of-mass frequencies and longitudinal
breathing frequencies vs particle number $n$ determined from thermally
excited oscillations in (a) the 25.4 mm $\times$ 50.8 mm confining
well and (b) the 14.0 mm $\times$ 50.8 mm well. Broken lines are
average values, and the solid line in (b) is a linear fit to $\omega_{br}$.}

\end{figure}

The anisotropy parameter $\alpha^{2}$ for each confining well was
determined from measurements of the center-of-mass frequencies excited
by thermal noise \cite{tes3} for the $x$ and $y$ directions, as
shown in Fig. \ref{fig:wxwy}. For the 25.4 mm $\times$ 50.8 mm well
{[}Fig. \ref{fig:wxwy}(a)], mode temperatures were found to be $300-400\;{\rm K}$,
indicating that the clusters are stable and in equilibrium with the
neutral gas component. The c.m. frequencies do not depend on the number
of particles, so the clusters do not perturb the potential well. By
averaging over the measured c.m. frequencies, we find $\omega_{0x}=7.00\pm0.06\;{\rm rad/s}$
and $\omega_{0y}=21.2\pm0.1\;{\rm rad/s}$. Equation (\ref{eq:nondim_para})
then gives $\alpha^{2}=9.24\pm0.2$, so that the anisotropy parameter
has been precisely determined.

For the 14.0 mm $\times$ 50.8 mm well {[}Fig. \ref{fig:wxwy}(b)],
data were taken at $n=6$ for two different sets of particles and
$n=9$ with three different sets to estimate the spread in the measured
frequencies, as can be seen in Fig. \ref{fig:wxwy}(b). In comparison
to $d=25.4\;{\rm mm}$, we observed a larger range of mode temperatures,
$300-500\;{\rm K}$, indicating that the clusters are somewhat less
stable. The c.m. frequencies are again found to be independent of
$n$, and the average c.m. frequencies are $\omega_{0x}=5.37\pm0.06\;{\rm rad/s}$
and $\omega_{0y}=29.7\pm0.1\;{\rm rad/s}$, giving $\alpha^{2}=30.7\pm0.7$.
In comparison to the $d=25.4\;{\rm mm}$ case, $\omega_{0y}$ has
increased as expected, while $\omega_{0x}$ has decreased slightly,
even though the long side of the rectangular well (50.8 mm) has not
changed. This indicates that decreasing $d$ is pushing the sheath
out of the concave depression formed by the bars. 

The Debye shielding parameter $\kappa$ was estimated by comparing
measured longitudinal breathing frequencies for several of the linear
configurations to normal mode frequencies calculated using the model.
Since the breathing oscillation varies the interparticle spacing,
it probes the dependence of the interparticle potential on particle
separation, and therefore $\kappa$. For an unshielded Coulomb interaction
$\kappa=0$, the squared normalized breathing frequency $\left(\omega_{br}/\omega_{0x}\right)^{2}=3$
irrespective of $n$, and the unshielded Coulomb regime is $\kappa\lesssim0.2$.
For the 25.4 mm $\times$ 50.8 mm well with $n=4$, the experimental
value $\omega_{br}=14.08\;{\rm rad/s}$, so that $\left(\omega_{br}/\omega_{0x}\right)^{2}=4.05$,
giving $\kappa\approx2.6$, while for $n=5$, the experimental value
$\omega_{br}=14.27\;{\rm rad/s}$, so that $\left(\omega_{br}/\omega_{0x}\right)^{2}=4.16$,
giving $\kappa\approx3.1$. The measured values of $\left(\omega_{br}/\omega_{0x}\right)^{2}$
are clearly not consistent with $\kappa=0$, and we conclude that
Debye shielding cannot be neglected when modeling these clusters.
The uncertainty in $\kappa$ is fairly large, so we take $\kappa=3.0$.
Using this value of $\kappa$, we compare the measured cluster width
$y_{rms}$ to the dimensionless cluster width to find $r_{0}=1.40\;{\rm mm}$,
$q=-1.3\times10^{4}e$ and $\lambda=0.47\;{\rm mm}$. These values
of $q$ and $\lambda$ are consistent with measurements made in isotropic
wells for similar discharge conditions \cite{tes1,tes2, tes3, tes4, tes5}.
In this case, the particle separation, which is $a=0.72\;{\rm mm}$
at the center of the $n=5$ cluster, is greater than the Debye length,
emphasizing the importance of Debye shielding. 

For $d=14.0\;{\rm mm}$ the Debye shielding parameter was estimated
from the normalized breathing frequencies for clusters with $n=6,$
8 and 9 particles. For $n=6$ we find $\left(\omega_{br}/\omega_{0x}\right)^{2}=4.32$
and 4.38, for $n=8$, $\left(\omega_{br}/\omega_{0x}\right)^{2}=4.20$
and for $n=9$, $\left(\omega_{br}/\omega_{0x}\right)^{2}=3.98$,
4.24 and 4.78. From this data we estimate $\kappa\approx4$, giving
$r_{0}=1.65\;{\rm mm}$, $q=-1.3\times10^{4}e$ and $\lambda=0.41\;{\rm mm}$.
As we show in the next section, $\kappa=4$ is very close to the critical
value for the zigzag transition, which may somewhat explain the spread
in the breathing frequencies for $n=9$. The physical parameters $q$
and $\lambda$ are consistent with the values found for the $d=25.4\;{\rm mm}$
well even though $\kappa$ is somewhat larger due to the decrease
in $\omega_{0x}$ {[}Eq. (\ref{eq:ka2})].

Equilibrium configurations computed from the model {[}Eq. (\ref{eq:nondim_pot})]
for $\alpha^{2}=9.24$ and $\kappa=3$ are shown in Figs. \ref{fig:d25pos}(h)-(n)
for comparison to the experimental configurations. For each value
of $n$, the experimental and predicted positions are very similar,
and the particle arrangements are identical. In particular, the zigzag
transition occurs at $n=6$ in both cases, so that experimentally
the critical value $n_{c}=6$. For $\alpha^{2}=9.24$ and $\kappa=0$,
$n_{c}=7$, which does not agree with the experimental results. A
comparison between the measured configurations and configurations
computed for $\alpha^{2}=30.7$ and $\kappa=4$ is shown in Fig. \ref{fig:d14pos}.
Again, the measured and computed configurations show excellent agreement
and the particle arrangements are identical. Further, the critical
value of $n$ for the zigzag transition is the same for the experimental
and the model results. For $\alpha^{2}=30.7$ and $\kappa=0$, the
critical value $n_{c}=13$. In the experiment we find $n_{c}=10$,
so that our results are not consistent with $\kappa=0$. The very
good agreement between experiment and model, and the consistency of
the results for two different confining wells, indicates that the
experimental results are robust.

\begin{figure}
\includegraphics[width=3.25in]{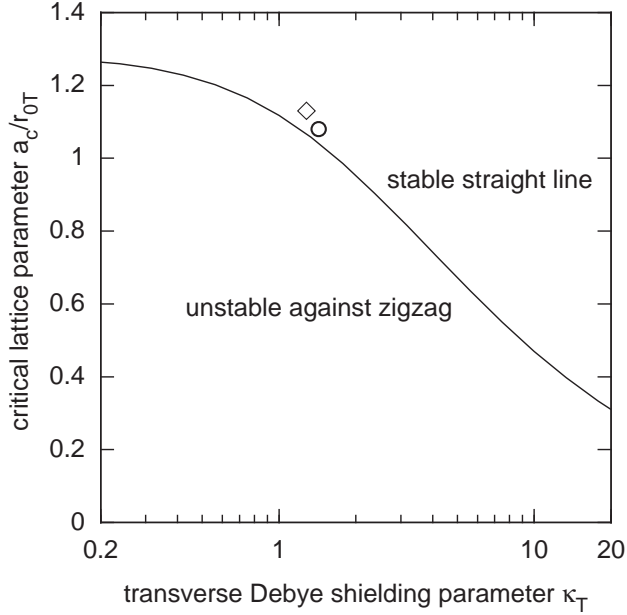}

\caption{\label{fig:acrit}Critical lattice parameter $a_{c}/r_{0T}$ for an
unbounded straight chain {[}Eq. (\ref{eq:ac})] vs the transverse
Debye shielding parameter $\kappa_{T}$. The data points are the experimentally
measured values for the last straight configuration for $d=25.4\;{\rm mm}$
(circle), and $d=14.0\;{\rm mm}$ (diamond). The measured points lie
close to, but above, the stability curve in the stable region.}

\end{figure}

In Fig. \ref{fig:acrit} we compare the values of the experimentally
measured lattice constant $a$ for the last straight configurations,
$n=n_{c}-1$, to the unbounded theory of Eq. (\ref{eq:ac}) for the
critical lattice constant. For both potential wells, the last straight
configuration has $n$ odd, so we approximate $a$ by the average
of the distances between the central particle and its two nearest
neighbors. For $d=25.4\;{\rm mm}$ with $n=5$, we find $a=0.72\;{\rm mm}$,
and for $d=14.0\;{\rm mm}$ with $n=9$, we find $a=0.60\;{\rm mm}$.
Using the measured values of $r_{0}$, $\kappa$ and $\alpha^{2}$
we then calculate for $d=25.4\;{\rm mm}$: $a/r_{0T}=1.08$ and $\kappa_{T}=1.43$,
and for $d=14.0\;{\rm mm}$: $a/r_{0T}=1.13$ and $\kappa_{T}=1.28$.
The experimental points lie close to the instability line, but slightly
above it, in the stable region. Since for both cases $a/r_{0T}<1.28$,
decreasing $\kappa_{T}$ (e.g., increasing $\lambda$) while holding
$a/r_{0T}$ constant moves the cluster into the unstable region, causing
a zigzag transition. Here finite size effects do not appear to be
very important, which may be because for $\kappa_{T}\gtrsim1$ the
zigzag instability is dominated by nearest neighbor interactions \cite{tesp}.

\begin{figure}
\includegraphics[width=3.25in]{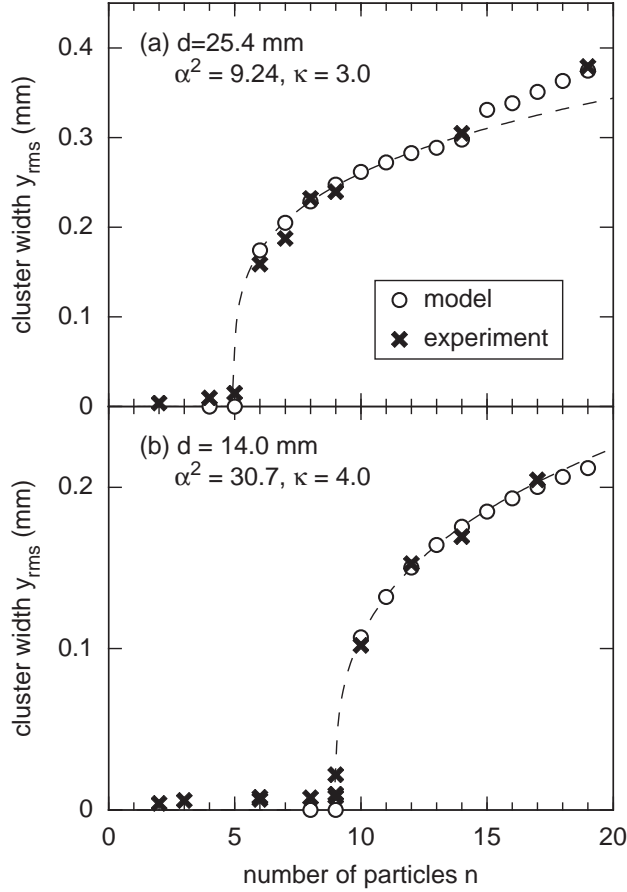}

\caption{\label{fig:yrmsvsn}Dependence of cluster width $y_{rms}$ on particle
number $n$ comparing experiment and model. (a) Experimental data
for bar separation $d=25.4\;{\rm mm}$ and model solutions with $\alpha^{2}=9.24$
and $\kappa=3$ scaled using $r_{0}=1.65\;{\rm mm}$. (b) Experimental
data for bar separation $d=14.0\;{\rm mm}$ and model solutions with
$\alpha^{2}=30.7$ and $\kappa=4$ scaled using $r_{0}=1.40\;{\rm mm}$.
The dashed lines are power law fits to the model points. }

\end{figure}

In Fig. \ref{fig:yrmsvsn} we compare the measured cluster width $y_{rms}$
with model solutions as a function of $n$ for the measured values
of $\alpha^{2}$ and $\kappa$. The only adjustable parameter is the
length scale for the cluster $r_{0}$, which was chosen to give good
agreement between the model and experiment. In both cases we see an
abrupt increase in the cluster width which is associated with the
zigzag transition. Above the transition, the data exhibit a power
law behavior which is consistent with a continuous phase transition.
Agreement between the model and the experiment is quite good. For
$d=25.4\;{\rm mm}$ there is a second structural transition at $n=15$
which corresponds to the change from a zigzag configuration to an
elliptical shell configuration {[}Fig. \ref{fig:d25pos}(g)] \cite{can,tes4}.
This transition may be roughly analogous to the transition from two
to three parallel chains in the unbounded system \cite{pia}. Such
a transition is not seen for $d=14.0\;{\rm mm}$ since $n$ is not
large enough, as confirmed by the fact that the $n=17$ configuration
is an extended zigzag {[}Fig. \ref{fig:d14pos}(g)].

\section{Dimensional Phase Transitions}

A phase transition is a sudden change in some property of a system,
called an order parameter, due to a small change in a control parameter.
Within this conceptual framework, the zigzag transition in these clusters
can be viewed as a dimensional phase transition between one-dimensional
and two-dimensional states \cite{sch}. We characterize the cluster
size in the longitudinal ($x$) and transverse ($y$) directions by
the rms values of the particle positions in the respective directions
{[}Eq. (\ref{eq:yrms})]. In particular, $y_{rms}$ is a good choice
for an (unnormalized) order parameter since $y_{rms}=0$ in the 1D
configuration and $y_{rms}>0$ in the 2D (zigzag) configuration. The
state variables that determine the system configuration are then $n$,
$\kappa$ and $\alpha^{2}$, where $n$ is discrete and $\kappa$
and $\alpha^{2}$ are continuous.

\begin{figure}
\includegraphics[width=3.25in]{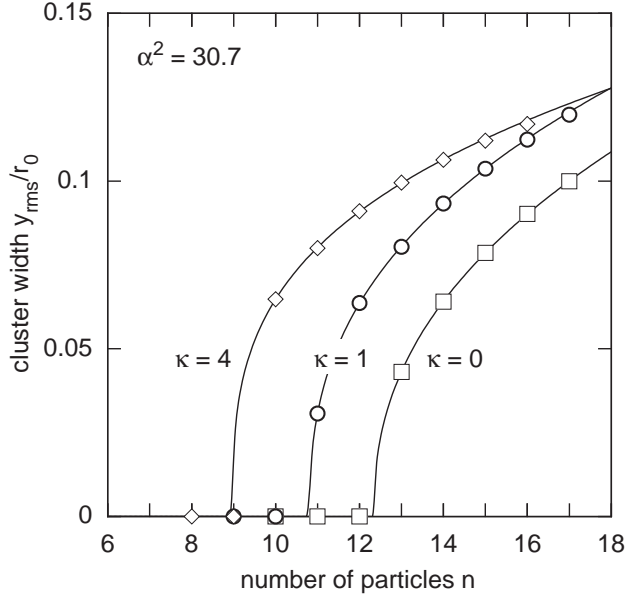}

\caption{\label{fig:yrmsvsnvsk}Cluster width $y_{rms}/r_{0}$ vs particle
number $n$ for Debye shielding parameters $\kappa=0$, 1 and 4 with
well anisotropy $\alpha^{2}=30.7$. The critical value of $n$ at
which the zigzag transition occurs decreases as $\kappa$ increases.
Solid lines are power law curves fitted to the first five points following
the zigzag transition.}

\end{figure}

Figure \ref{fig:yrmsvsnvsk} demonstrates that $n_{c}$ decreases
as $\kappa$ increases using $\alpha^{2}=30.7$. That is, expressions
which predict the critical value of $\alpha^{2}$ for a given $n$
with $\kappa=0$ \cite{can} are incorrect when the strength of Debye
shielding is such that the interaction is not essentially unshielded.
Interestingly, even though $n$ is small and discrete, the cluster
width above the transition is well characterized by a power law \cite{pia}\begin{equation}
y_{rms}\propto\left(n-n_{c}'\right)^{\nu}\label{eq:ncp}\end{equation}
where $n_{c}'$ is a continuous critical $n$, and $n_{c}=\left\lceil n_{c}'\right\rceil $.
That $n_{c}'$ is continuous indicates there may be a continuum theory
for the zigzag transition where $n$ is also continuous. For $\alpha^{2}=30.7$
and $\kappa=0,$1 and 4, we fit the first five points after the transition
to find $n_{c}'=12.34$, 10.82 and 8.99, and a critical exponent $\nu=0.430$,
0.387 and 0.310, respectively. Here $\nu$ decreases with increasing
$\kappa$. As discussed above, $\kappa=4$ is very close to the critical
value when $\alpha^{2}=30.7$, and we find $n_{c}'$ very close to
an integer value. In fact, for $\kappa=4$, $n_{c}=10$ so we expect
$n_{c}'>9$, which is not satisfied here due to a small uncertainty
in the fitting coefficients.

\begin{figure}
\includegraphics[width=3.25in]{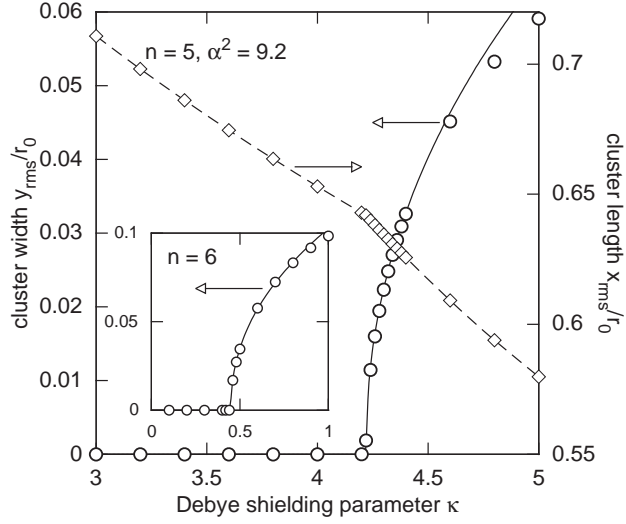}

\caption{\label{fig:n=00003D5transition}Computed cluster width and length
for anisotropy parameter $\alpha^{2}=9.2$ for $n=5$ and (inset)
$n=6$ particles vs the Debye shielding parameter $\kappa$. Solid
lines show the power law fit to Eq. (\ref{eq:kcrit}), while the dashed
line has been added to guide the eye. For $n=5$ particles there is
a critical value $\kappa_{c}=4.22$ below which the cluster is one
dimensional and above which it is two dimensional. For $n=6$, $\kappa_{c}=0.45$.}

\end{figure}

The computed dependence of cluster length and width for $n=5$ and
6 and $\alpha^{2}=9.2$ on the Debye shielding parameter $\kappa$
is shown in Fig. \ref{fig:n=00003D5transition}. For the finite model,
in contrast to the unbounded case \cite{pia}, increasing $\kappa$
decreases the nearest neighbor distance, and therefore the linear
particle density. For $n=5$ the critical value of $\kappa$ for the
zigzag transition is $\kappa_{c}=4.22$. That is, the cluster is in
a 1D configuration for $\kappa<4.22$. For $\kappa>4.22$, $y_{rms}$
is positive and increases rapidly with $\kappa$. The cluster length
$x_{rms}$ has a discontinuous first derivative at the phase transition.
The inset of Fig. \ref{fig:n=00003D5transition} shows the transition
for $n=6$, where the critical value is $\kappa_{c}=0.45$. Consequently,
for $n=6$ and $\kappa<0.45$ the cluster is linear. Since we find
experimentally that the $n=6$ cluster is in the zigzag configuration,
we conclude that the experimental Debye shielding parameter \emph{must}
lie in the interval $0.45<\kappa<4.22$. For $\kappa>\kappa_{c}$,
the cluster width has a power law behavior\begin{equation}
y_{rms}\propto\left(\kappa-\kappa_{c}\right)^{\beta},\label{eq:kcrit}\end{equation}
where $\beta$ is a critical exponent that is independent of the normalization
of $y_{rms}$. For the $n=5$ and 6 cases illustrated in Fig. \ref{fig:n=00003D5transition},
we find $\beta=0.463$ and 0.450, respectively. This analysis was
repeated for $\alpha^{2}=30.7$ with $n=9$ and 10. For $n=9$, $\kappa_{c}=4.08$
and $\beta=0.463$, while for $n=10$, $\kappa_{c}=1.77$ and $\beta=0.469$.
Experimentally, we find $n_{c}=10$, which means that for the experiment
$\kappa$ must lie in the interval $1.77<\kappa<4.08$. The critical
exponent for the zigzag transition vs $\kappa$ is nearly the same
for the four cases considered here, so it may be that in Yukawa clusters
there is a universal critical exponent $\beta\approx0.46$ for the
zigzag transition caused by changing the Debye shielding parameter
$\kappa$.

\begin{figure}
\includegraphics[width=3.25in]{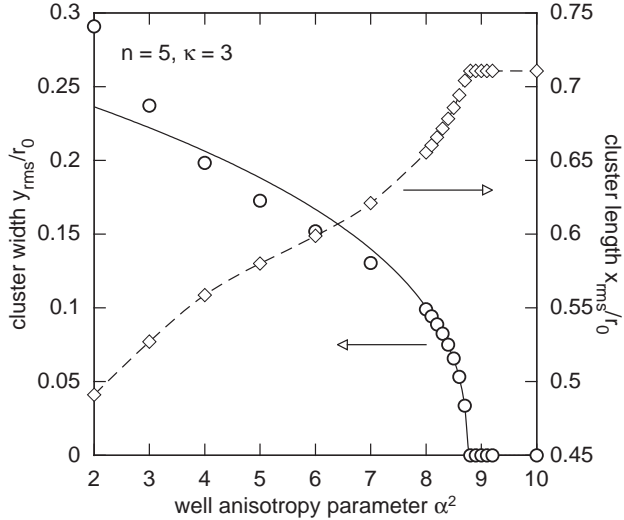}

\caption{\label{fig:k=00003D3transition}Computed cluster width and length
for $n=5$ with shielding parameter $\kappa=3$ vs anisotropy parameter
$\alpha^{2}$. A 2D-1D phase transition is with a critical value $\alpha_{c}^{2}=8.74$.
The fitted power law (solid line) gives a critical exponent $\gamma=0.387$.}

\end{figure}

For fixed values of $n$ and $\kappa$, a 2D-1D transition (an {}``inverse
zigzag'') takes place as the anisotropy parameter $\alpha^{2}$ increases,
as shown in Fig. \ref{fig:k=00003D3transition} for $n=5$ and $\kappa=3$.
As $ $$\alpha^{2}$ increases, the dimensionless cluster length increases
while the width decreases. Near the transition, the width exhibits
a power-law approach to $y_{rms}=0$. Assuming\begin{equation}
y_{rms}\propto\left(\alpha_{c}^{2}-\alpha^{2}\right)^{\gamma},\label{eq:a2crit}\end{equation}
we find the critical value $\alpha_{c}^{2}=8.74$ and the critical
exponent $\gamma=0.387$. That is, if $\kappa=3$ and $\alpha^{2}<8.74$
then $n=5$ particles will be in a zigzag configuration. $ $These
results are consistent with the experiment where we observed a 1D
configuration for $\alpha^{2}=9.24$. When $\alpha^{2}>\alpha_{c}^{2}$
the model results are independent of $\alpha^{2}$, which can be seen
by the constancy of $x_{rms}$, since $y_{i}=\eta_{i}=0$ and the
configuration only depends on $n$ and $\kappa$.

\section{Conclusions}

We have studied one- and two-dimensional Yukawa clusters with a small
number of particles $n\le19$ confined in biharmonic potential wells
both experimentally and theoretically. Experiments were performed
in the Dusty Ohio Northern University experimenT (DONUT). For $n$
less than a critical value $n_{c}$, the clusters are in a one-dimensional
straight line state. When $n=n_{c}$ the cluster undergoes a zigzag
transformation to a two-dimensional state. In our experiments, the
anisotropy of the confining potential well was accurately determined
by measuring the frequencies of center-of-mass oscillations excited
by thermal noise in both the $x$ (longitudinal) and $y$ (transverse)
directions and the Debye shielding parameter was estimated from the
measured longitudinal breathing frequency of 1D clusters. Experimental
and model data show excellent quantitative agreement, confirming that
dusty plasma is a very good real-world system for studying 1D and
2D Yukawa systems and the transitions between these states. In particular,
strongly-coupled linear configurations with $n<n_{c}$ are true 1D
systems for which normal modes are either purely longitudinal or purely
transverse.

Our results clearly show that Debye shielding is important for our
experimental conditions, and our results are not consistent with physics
in the unshielded Coulomb regime. For a given value of the potential
well anisotropy, the critical particle number $n_{c}$ decreases as
the shielding parameter $\kappa$ increases (i.e., as the Debye length
decreases). For the measured well anisotropies, the experimental values
of $n_{c}$ are below those predicted for an unshielded Coulomb interaction
($\kappa=0$), indicating $\kappa>0$. This is reinforced by noting
that the measured Debye lengths are less than the particle separation
and that the normalized squared longitudinal breathing frequencies
$\left(\omega_{br}/\omega_{0x}\right)^{2}>3$.

Our finding that $\kappa>0$ contradicts the conclusion in Ref. \cite{mel1}
for a similar experiment where the results where said to be consistent
with an unshielded Coulomb interaction. The method used in Ref. \cite{mel1}
to determine cluster parameters is a static analysis that treats both
the anisotropy parameter $\alpha^{2}$ and shielding parameter $\kappa$
as free parameters and compares the observed $n_{c}$ with that predicted
for $\kappa=0$. When the decrease in $n_{c}$ with $\kappa$ described
in the present work is considered, it seems likely that static analysis
method \cite{mel1} is only weakly constrained and cannot be used
to accurately determine cluster parameters. 

Finally, we have demonstrated that the zigzag transition in a Yukawa
cluster can be viewed as a phase transition from a one-dimensional
state to a two-dimensional state. Though this was previously demonstrated
for unbounded systems \cite{pia}, here the number of particles is
finite and really quite small. This is true for transitions initiated
by changing the Debye shielding parameter, the potential well anisotropy
and the number of particles. In all three cases, we find that the
transverse cluster width has a power law dependence near the transition,
indicated that transition behaves as a continuous phase transition
with a critical exponent. For 1D-2D transitions caused by increasing
the Debye shielding parameter, we tentatively identify a universal
critical exponent {[}Eq. (\ref{eq:kcrit})] with a value $\beta\approx0.46$.

\begin{acknowledgments}
T.E.S. would like to thank Ohio Northern University for sabbatical
release time.
\end{acknowledgments}

\end{document}